\documentstyle[12pt]{article}

\begin{document}

\title{A different see-saw formula for neutrino masses}

\author{{\bf S.M. Barr}\\Bartol Research Institute\\
University of Delaware\\Newark, DE 19716}

\date{}
\maketitle

\begin{abstract}

In a wide class of unified models there is an additional (and 
possibly dominant) term in the neutrino mass formula that under the 
simplest assumption takes the form $M_{\nu} = (M_N + M_N^T)u/M_G$, 
where $M_N$ is the neutrino Dirac mass matrix, and $u = O(M_W)$.
This makes possible highly predictive models. A generalization of this 
form yields realistic neutrino masses and mixings more readily than the 
usual see-saw formula in some models.

\end{abstract}

In grand unified theories based on $SO(10)$ or larger groups, right-handed
neutrinos exist and typically acquire mass of order the unification
scale, $M_G \simeq 2 \times 10^{16}$ GeV. When these superheavy 
neutrinos are integrated out, masses are induced for the left-handed neutrinos
that are of order $v^2/M_G \simeq 10^{-3}$ eV, where $v = \sqrt{v_u^2 +
v_d^2} = (\sqrt{2} G_F)^{-1/2}$ is the weak scale. This is the so-called
see-saw mechanism \cite{seesaw}, and it well explains --- at least in an
order of magnitude sense --- the scale of neutrino masses seen in atmospheric
and solar neutrino oscillations ($\sqrt{\delta m^2_{atm}} \simeq 5 \times
10^{-2}$ eV \cite{atm} and $\sqrt{\delta m^2_{sol}} \simeq 8.5 \times
10^{-3}$ eV \cite{sol}). 

It has been long known \cite{type2} that in certain unified schemes 
$SU(2)_L$-triplet Higgs fields exist that couple directly to the left-handed
neutrinos and give them a Majorana mass of order $v^2/M_G$. This effect is
called the type II see-saw mechanism. It typically happens, for example,
in $SO(10)$ models in which a $\overline{{\bf 126}}$ and ${\bf 126}$ of
Higgs fields are responsible for breaking $B-L$ and generating the right-handed
neutrino mass matrix, $M_R$. In such models, both the type I (i.e. original)
and the type II see-saw mechanisms would normally operate.

In this letter we point out that in $SO(10)$ models where a 
$\overline{{\bf 16}}$ and ${\bf 16}$ of Higgs fields (rather than
$\overline{{\bf 126}}$ and ${\bf 126}$) are responsible for breaking $B-L$ 
and generating $M_R$ another type of see-saw mechanism operates that we
shall call type III. The type I see-saw formula is

\begin{equation}
M_{\nu}^{I} = - M_N M_R^{-1} M_N^T,
\end{equation}

\noindent
where $(M_N)_{ij} \nu_i N^c_j = (Y_N)_{ij} \nu_i N^c_j \langle H_u \rangle$
is the Dirac neutrino mass term and $(M_R)_{ij} N^c_i N^c_j$ is the 
Majorana mass term of the right-handed neutrinos. Eq. (1) can be understood 
diagrammatically as arising from the graph in Fig. 1. 

The type II see-saw formula is simply
$M_{\nu}^{II} = M_T$,
where $(M_T)_{ij} \nu_i \nu_j = (Y_T)_{ij} \nu_i \nu_j \langle T \rangle$
is the direct Majorana mass term of the left-handed neutrinos coming
from their coupling to the triplet Higgs field $T$. In $SO(10)$, if $M_R$
comes from the Yukawa term $(Y_R)_{ij} {\bf 16}_i {\bf 16}_j 
\overline{{\bf 126}}_H$, the same term will generate the matrix $M_T$, since
$\overline{{\bf 126}}_H$ contains the triplet $T$. In the very simplest 
$SO(10)$ models with one $\overline{{\bf 126}}_H$, therefore, one would expect
$M_{\nu}^{II} = M_T = c M_R (v/M_G)^2$,
where $c \sim 1$ \cite{type2}. The reason that the vacuum expectation 
value (VEV) of $T$ is of order $v^2/M_G$ is simple. If $T$ has a conjugate 
field $\overline{T}$, then one expects that in the superpotential there will be
terms of the form $M_T \overline{T} T + \overline{T} H_u H_u$, where 
$M_T \sim M_G$. Integrating out $\overline{T}$ gives $- F_T^* =
M_T \langle T \rangle + \langle H_u \rangle^2 = 0$. 

The type I and type II see-saw formulas can be understood as arising from 
block-diagonalizing the complete mass matrix of the neutrinos and 
anti-neutrinos:
\begin{equation}
{\cal L}_{\nu \; mass} = \left( \nu_i, N^c_i \right) \left(
\begin{array}{cc} (M_T)_{ij} & (M_N)_{ij} \\ (M_N^T)_{ij} & 
(M_R)_{ij} \end{array} \right)
\left( \begin{array}{c} \nu_j \\ N^c_j \end{array} \right),
\end{equation}

\noindent
with $M_R \sim M_G$, $M_N \sim v$, $M_T \sim v^2/M_G$, giving
$M_{\nu} = M_{\nu}^I + M_{\nu}^{II} = - M_N M_R^{-1} M_N^T + M_T$, neglecting
terms higher order in $v/M_G$.

In the alternative where $B-L$ breaking is produced by the expectation 
values of $\overline{{\bf 16}}$ and ${\bf 16}$
of Higgs fields, $M_R$ arises from an effective operator of the form
\begin{equation}
{\cal O}_R = (Y_R)_{ij} {\bf 16}_i {\bf 16}_j \overline{{\bf 16}}_H 
\overline{{\bf 16}}_H/M_G.
\end{equation}

\noindent
Such an operator comes from the diagram in Fig. 2. The VEV $\langle 
{\bf 1}(\overline{{\bf 16}}_H) \rangle \equiv \Omega$ and the mass matrix
$M_{mn}$ in that diagram are $O(M_G)$. ${\bf M}({\bf N})$ denotes 
an $SU(5)$ ${\bf M}$ representation contained
in an $SO(10)$ ${\bf N}$. We assume for simplicity that the fields 
$S_m$ in Fig. 2 are singlets under $SO(10)$ as well as $SU(5)$, though it would
change nothing in the later discussion if they were in other
representations of $SO(10)$.

In addition to the couplings $M_{mn} {\bf 1}_m {\bf 1}_n$ and 
$F_{im} {\bf 16}_i {\bf 1}_m \overline{{\bf 16}}_H$ involved in Fig. 2,
there is the Dirac mass term $(M_N)_{ij} \nu_i N^c_j$ that comes
from Yukawa terms of the form ${\bf 16}_i {\bf 16}_j H$, where $H \subset 
\overline{{\bf 16}} \times \overline{{\bf 16}}$. 
The full mass matrix of the neutrinos and 
anti-neutrinos (which now include $\nu_i$, $N^c_i$, and $S_m$) is
\begin{equation}
{\cal L}_{\nu \; mass} = \left( \nu_i, N^c_i, S_m \right) \left(
\begin{array}{ccc} 0 & (M_N)_{ij} & 0 \\ (M_N^T)_{ij} & 0 & 
F_{in} \Omega \\ 0 & F^T_{mj} \Omega & M_{mn} \end{array} \right)
\left( \begin{array}{c} \nu_j \\ N^c_j \\ S_n \end{array} \right),
\end{equation}

\noindent
where $i,j = 1,2,3$ and $m,n = 1, ... , N$, and $N$ is the number
of species of singlets $S_m$. It is easy to show that the effective
mass matrix $M_{\nu}$ of the light neutrinos is given, up to negligible
corrections higher order in $v/M_G$, by
\begin{equation}
M_{\nu} = - M_N (F \Omega \; M^{-1} \; F^T \Omega)^{-1} M_N^T.
\end{equation}

\noindent
In other words, one has the usual type I see-saw formula with
\begin{equation}
M_R = (F \Omega) M^{-1} (F^T \Omega).
\end{equation}

In such a model there is no type II see-saw contribution, as the 
$\overline{{\bf 16}}$ and ${\bf 16}$ do not contain a weak-triplet Higgs
field. However, another type of new see-saw contribution can 
arise as we will now see.. 
So far, we have only taken into account the VEV of the
${\bf 1}(\overline{{\bf 16}})$ component of the $\overline{{\bf 16}}_H$, which
we called $\Omega$. However, there is a weak doublet in the 
${\bf 5}(\overline{{\bf 16}})$ that can have a weak-scale VEV, which we shall
call $u$. There is no a priori reason why $u$ should vanish. If it does
not, then the term $F_{im} {\bf 16}_i {\bf 1}_m \overline{{\bf 16}}_H$ not only
produces the $O(M_G)$ mass term $F_{im}(N^c_i S_m) \Omega$, but an
$O(v)$ mass term $F_{im}(\nu_i S_m) u$. Eq. (4)
then becomes
\begin{equation}
{\cal L}_{\nu \; mass} = \left( \nu_i, N^c_i, S_m \right) \left(
\begin{array}{ccc} 0 & (M_N)_{ij} & F_{in} u \\ (M_N^T)_{ij} & 0 & 
F_{in} \Omega \\ F^T_{mj} u & F^T_{mj} \Omega & M_{mn} \end{array} \right)
\left( \begin{array}{c} \nu_j \\ N^c_j \\ S_n \end{array} \right),
\end{equation}

\noindent
This can be simplified by a rotation in the $\nu_i N^c_i$ plane by angle 
$\tan^{-1}(u/\Omega)$: 
$\nu'_i = (\nu_i - \frac{u}{\Omega} N^c_i)/\sqrt{1 + (u/\Omega)^2}$,
$N^{c \prime}_i = (N^c_i + \frac{u}{\Omega} \nu_i)/\sqrt{1 + (u/\Omega)^2}$,
which has the effect of eliminating the $\nu S$ entries. 
It also replaces the $0$ in the 
$\nu \nu$ entry by 
\begin{equation}
M_{\nu}^{III} = - (M_N + M_N^T) \frac{u}{\Omega},
\end{equation}

\noindent
neglecting, as always, terms higher order in $v/M_G$.
Otherwise, the resulting matrix has the same form as Eq. (4). Therefore, 
the full result for $M_{\nu}$ is given by the sum of Eqs. (5) and (8).

The relative size of the two contributions to $M_{\nu}$ is model dependent.
Since $M_N$ is related to the up quark mass matrix $M_U$ by $SO(10)$, 
one would expect the entries for the first and second families to be very small
compared to $v$. Consequently, due to the fact that $M_N$ comes in squared 
in $M_{\nu}^I$ but only linearly in $M_{\nu}^{III}$, the latter should 
dominate, except perhaps for the third family. 
$M_{\nu}^{III}$ would also dominate if the elements
of $M_{mn}$ were small compared to $\Omega \simeq M_G$, as Eqs. (5) and (6) 
show.

That $M_{\nu}^{III}$ dominates is an interesting possibility, as
remarkably predictive 
$SO(10)$ models of quark and lepton masses would then be constructable. 
Usually the most one can achieve in models where $M_{\nu}$ is given by
the type I see-saw formula is predictions for the mass matrices of the 
up quarks, down quarks, and charged leptons ($M_U$, $M_D$, $M_L$), and for
the Dirac mass matrix of the neutrinos ($M_N$), since these four matrices are 
intimately related to each other by symmetry. (For example in the ``minimal
$SO(10)$ model" they all come from one term $Y_{ij} {\bf 16}_i {\bf 16}_j
{\bf 10}_H$ and have exactly the same form.) However, sharp predictions for
neutrino masses and mixings are hard to achieve because of the difficulty in
constraining the form of $M_R$, which comes from different terms. On the other
hand, if the type III see-saw contributions are dominant, then the 
matrix $M_R$ is irrelevant; a knowledge of $M_N$ and $M_L$ is sufficient to
determine the neutrino mass ratios and mixing angles.

In this letter we will not be so ambitious. Rather we will look at a version 
of the type III see-saw that is less predictive but still has certain
attractive features. In the foregoing, we assumed that there was only a single
$\overline{{\bf 16}}$ of Higgs fields that contributed to neutrino masses.
If there is more than one, then their coupling to neutrinos comes from the
term $\sum_{aim} F_{im}^a ({\bf 16}_i {\bf 1}_m) \overline{{\bf 16}}_{Ha}$,
which contains $\sum_{im} F_{im} (N^c_i S_m) \Omega + 
\sum_{im} F'_{im} (\nu_i S_m) u$, where $F_{im} \equiv \sum_a F_{im}^a
\Omega_a/\Omega$, $F'_{im} \equiv \sum_a F_{im}^a u_a/u$, $\Omega \equiv
(\sum_a \Omega_a^2)^{1/2}$, and $u \equiv
(\sum_a u_a^2)^{1/2}$. Then Eq. (7) is modified by having 
Yukawa matrices in the $\nu S$ and $N^c S$ blocks that are no longer
proportional to each other. It is then not possible
to null out the $\nu S$ block of $M_{\nu}$ by a simple flavor-independent
rotation by angle $\tan^{-1} (u/\Omega)$, as in the special case discussed
above. Consequently, the effective light-neutrino mass matrix is more
complicated. In the most general case it can be
written, $M_{\nu} = -\tilde{M}_N M_R^{-1} \tilde{M}_N^T +
(F' u) M^{-1} (F^{\prime T} u)$, where $M_R$ is given by Eq. (6) as before, 
and $\tilde{M}_N \equiv M_N + (F' u) M^{-1} (F^T \Omega)$. However, a
great simplification results if one assumes that the number of species of
singlet fermions $S_m$ is three, i.e. one for each family. (If there were
less than three, not all the $N^c_i$ would get superlarge mass, and some of the 
light neutrinos would have masses of order $v$.) With three species of
$S_m$, the matrices $F$ and $F'$ are (generally) invertible and then
$M_{\nu} = M_{\nu}^I + M_{\nu}^{III}$,
where $M_{\nu}^I$ is given as before by Eq. (5) and
\begin{equation}
M_{\nu}^{III} = - (M_N H + H^T M_N^T) \frac{u}{\Omega}, \;\;\; 
H \equiv (F' F^{-1})^T.
\end{equation}

\noindent
In this the generalized type III see-saw formula, the dimensionless 
$3 \times 3$ matrix $H$ introduces many unknown parameters, more indeed 
than does $M_R$ in the type I see-saw. However, as we shall now show by 
an example, in $SO(10)$ models it may be easier to obtain a realistic 
pattern of neutrino masses and mixings without fine-tuning of parameters 
in the generalized type III see-saw than in the type I see-saw.

The $SO(10)$ model of Ref. \cite{ab} gives an excellent fit to the quark 
masses and mixings and the charged lepton masses, fitting 13 real
quantities with 8 real parameters. This fit uniquely determines the
neutrino Dirac mass matrix (at the unification scale) to be
\begin{equation}
M_N = \left( \begin{array}{ccc} \eta & 0 & 0 \\ 0 & 0 & \epsilon \\
0 & -\epsilon & 1 \end{array} \right) m_U,
\end{equation}

\noindent
where $m_U \cong m_t$, $\eta \cong m_u^0/m_t^0 \cong 0.6 \times 10^{-5}$,
and $\epsilon \cong 3 \sqrt{m_c^0/m_t^0} \cong 0.14$. (Superscripts here refer
to quantities evaluated at $M_G$.) In this model there
is a very large (namely $\tan^{-1} 1.8$) contribution to the atmospheric
neutrino mixing angle coming from the charged lepton mass matrix $M_L$, which
is completely known. However, as $M_R$ is not known, it is impossible to 
predict the neutrino mass ratios and the other neutrino
mixing angles (or even the atmospheric angle precisely) within the framework
of the type I see-saw. Nevertheless, one can ask whether what we know about
these neutrino masses and mixings can be accomodated in the model
with a reasonable form for $M_R$. Parametrizing that matrix by
$(M_R^{-1})_{ij} = a_{ij} m_R^{-1} = a_{ji} m_R^{-1}$, 
the type I see-saw formula gives
\begin{equation}
M_{\nu} = \left( \begin{array}{ccc}
a_{11} \eta^2 & - a_{13} \epsilon \eta & (a_{13} + a_{12} \epsilon) \eta \\
- a_{13} \epsilon \eta & a_{33} \epsilon^2 & -(a_{33} 
+ a_{23} \epsilon) \epsilon \\ (a_{13} + a_{12} \epsilon) \eta & 
-(a_{33} + a_{23} \epsilon) \epsilon & a_{33} + 2 a_{23} \epsilon +
a_{22} \epsilon^2 \end{array} \right) m_U^2/m_R.
\end{equation}

\noindent
Neglecting the relatively small first row and column, the condition that
the ratio $m_2/m_3$ of the two heaviest neutrino masses be equal to some 
value $r$ is that
\begin{equation}
a_{22} a_{33} - a_{23}^2 \cong \frac{r}{(1+r^2)^2} (a_{33} \frac{1}{\epsilon^2}
+ 2a_{23} \frac{1}{\epsilon} + a_{22} + a_{33})^2.
\end{equation}

\noindent
It is evident that $r$ naturally is of order
$\epsilon^4 \approx 4 \times 10^{-4}$. For $r$ to be of order $\epsilon^0$ 
(as indicated by experiment, which gives $r \approx 1/6$) 
the elements must be somewhat ``tuned". For example,
setting $a_{23}/a_{33} = p \epsilon^{-1} + O(\epsilon^0)$ and 
$a_{22}/a_{33} = q \epsilon^{-2} + O(\epsilon^{-1})$, Eq. (12) gives
the condition $1 + 2p + q = 0$. In other words, not only must the 23 block
of $M_R$ have a hierarchy that is correlated with the hierarchy of the
23 block of $M_N$, but it must also satisfy a non-trivial numerical relation
among its elements. This kind of mild fine-tuning of the 23 block of
$M_R$ is typically required in $SO(10)$ models with the type 
I see-saw mechanism \cite{bd}

It can be seen from Eq. (11) that to fit the LMA solar solution  
$a_{11} \leq \epsilon^2/\eta^2$, $a_{12} \leq \epsilon/\eta$, and 
$a_{13} \sim \epsilon^2/\eta$. Thus the correlation between the hierarchies 
of $M_R$ and $M_N$ extends also to the first family.

By contrast, a satisfactory pattern of neutrino masses and mixings can 
be achieved without any fine-tuning in this model if the type III see-saw 
mechanism dominates. There are two interesting cases. Suppose, first, that all
the elements of $F$ are of the same order, and likewise for $F'$.
Then all the elements of $H \equiv (F' F^{-1})^T$ will be of order one.
>From Eq. (9), neglecting terms of order $\eta$,
\begin{equation}
M_{\nu} = \left( \begin{array}{ccc}
0 & \epsilon H_{31} & H_{31} - \epsilon H_{21} \\
\epsilon H_{31} & 2 \epsilon H_{32} & H_{32} + \epsilon(H_{33} - H_{22}) \\
H_{31} - \epsilon H_{21} & H_{32} + \epsilon(H_{33} - H_{22}) & 
2(H_{33} - \epsilon H_{23}) \end{array} \right) \frac{m_U u}{\Omega}.
\end{equation}

\noindent
Here it is clear that without any fine-tuning $|r|$ ($\equiv |m_2/m_3|$) is 
somewhat less than one, as desired. More precisely:
$-r/(1 + r^2) \cong \frac{1}{4} (H_{32}/H_{33})^2 + O(\epsilon)$.
Moreover, the LMA solution naturally emerges. For $U_{e3}$ to be 
consistent with present limits, $\epsilon H_{21}$ must approximately cancel
$H_{31}$ in the 13 and 31 elements of $M_{\nu}$. However, all the other 
elements of $H$ can be of order one.

Note that a satisfactory pattern of light neutrino masses and mixings 
emerges with {\it no hierarchy} among the 
superheavy neutrinos, which all have masses of order $M_G$, something 
that is impossible in the type I see-saw. This is an attractive possibility, 
but would create problems for leptogenesis \cite{leptogen}.

A second interesting case is that $F$ and $F'$ both have the form
\begin{equation}
F,F' \sim \left( \begin{array}{ccc} \eta/\epsilon & \eta/\epsilon &
\eta/\epsilon \\ 1 & 1 & 1 \\ 1 & 1 & 1 \end{array} \right), 
\end{equation}

\noindent
as might arise naturally if the first family had a different abelian family
charge than the others. Then, by Eq. (9), $M_{\nu}$ has the form
\begin{equation}
M_{\nu} \sim \left( \begin{array}{ccc} \eta & \epsilon & \epsilon \\
\epsilon & \epsilon & 1 \\ \epsilon & 1 & 1 \end{array} \right) 
\frac{m_U u}{\Omega},
\end{equation}

\noindent
that is, the same form as the previous case, except that $U_{e3}$ is 
automatically of order $\epsilon$. In this case, the superheavy neutrinos
consist of one pseudo-Dirac neutrino with mass $O((\eta/\epsilon) M_G)
\sim 10^{12}$ GeV and two pseudo-Dirac neutrinos with mass of 
order $M_G$.

\newpage

\noindent
{\bf\Large Figure Captions}

\vspace{1cm}

\noindent
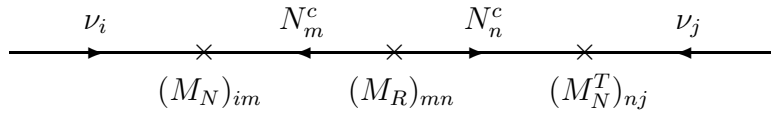
Fig. 1: Diagram that gives the light neutrinos type I see-saw masses of order
$v^2/M_G$.

\vspace{1cm}

\noindent
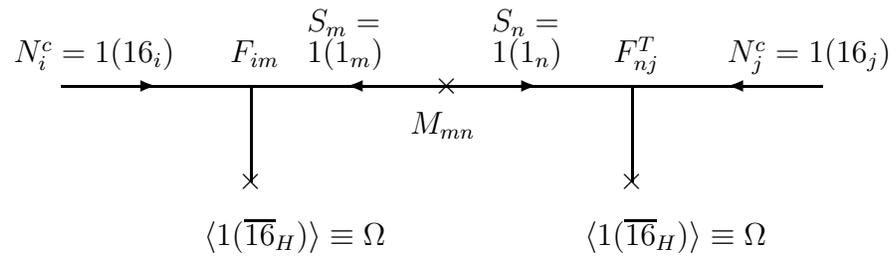
Fig. 2: Diagram that produces the effective operator 
${\bf 16}_i {\bf 16}_j \overline{{\bf 16}}_H \overline{{\bf 16}}_H/M_G$,
which generates $M_R$.

\newpage

\begin{picture}(360,100)
\thicklines
\put(36,68){\vector(1,0){36}}
\put(72,68){\line(1,0){36}}
\put(105,65){$\times$}
\put(108,68){\line(1,0){36}}
\put(180,68){\vector(-1,0){36}}
\put(177,65){$\times$}
\put(180,68){\vector(1,0){36}}
\put(216,68){\line(1,0){36}}
\put(249,65){$\times$}
\put(252,68){\line(1,0){36}}
\put(324,68){\vector(-1,0){36}}
\put(64,77){$\nu_i$}
\put(92,50){$(M_N)_{im}$}
\put(138,77){$N^c_m$}
\put(164,50){$(M_R)_{mn}$}
\put(208,77){$N^c_n$}
\put(240,50){$(M_N^T)_{nj}$}
\put(288,77){$\nu_j$}
\put(168,0){{\bf Fig. 1}}
\end{picture}

\vspace{2cm}

\begin{picture}(360,216)
\thicklines
\put(36,108){\vector(1,0){36}}
\put(72,108){\line(1,0){36}}
\put(108,72){\line(0,1){36}}
\put(103,69){$\times$}
\put(108,108){\line(1,0){36}}
\put(180,108){\vector(-1,0){36}}
\put(177,105){$\times$}
\put(180,108){\vector(1,0){36}}
\put(216,108){\line(1,0){36}}
\put(252,72){\line(0,1){36}}
\put(247,69){$\times$}
\put(252,108){\line(1,0){36}}
\put(324,108){\vector(-1,0){36}}
\put(18,117){$N^c_i=1(16_i)$}
\put(100,117){$F_{im}$}
\put(90,48){$\langle 1(\overline{16}_H) \rangle \equiv \Omega$}
\put(129,129){$S_m =$}
\put(129,117){$1(1_m)$}
\put(168,90){$M_{mn}$}
\put(199,129){$S_n=$}
\put(199,117){$1(1_n)$}
\put(245,117){$F^T_{nj}$}
\put(234,48){$\langle 1(\overline{16}_H)\rangle \equiv \Omega$}
\put(288,117){$N^c_j = 1(16_j)$}
\put(168,0){{\bf Fig. 2}}
\end{picture}

\end{document}